\documentstyle[preprint,aps]{revtex}

\begin{document}
\normalsize
\title{Energy and Charged Particle Flow in
        10.8 A GeV/c Au+Au Collisions}
\author{J.~Barrette$^4$, R.~Bellwied$^{8}$, S.~Bennett$^{8}$,
P.~Braun-Munzinger$^6$, W.~C.~Chang$^6$, W.~E.~Cleland$^5$,
M.~Clemen$^5$, J.~Cole$^3$, 
T.~M.~Cormier$^{8}$, G.~David$^1$, J.~Dee$^6$, O.~Dietzsch$^7$,
M.~Drigert$^3$, J.~R.~Hall$^8$, T.~K.~Hemmick$^6$,
N.~Herrmann$^2$, B.~Hong$^6$, Y.~Kwon$^6$,
R.~Lacasse$^4$, A.~Lukaszew$^8$, Q.~Li$^{8}$, T.~W.~Ludlam$^1$,
S.~K.~Mark$^4$, R.~Matheus$^{8}$, S.~McCorkle$^1$, J.~T.~Murgatroyd$^8$,
E.~O'Brien$^1$,  
S.~Panitkin$^6$, T.~Piazza$^6$, C.~Pruneau$^8$, M.~N.~Rao$^6$,
M.~Rosati$^4$, N.~C.~daSilva$^7$, S.~Sedykh$^6$, U.~Sonnadara$^5$,
J.~Stachel$^6$, E.~M.~Takagui$^7$, 
S.~Voloshin$^5\footnote{On leave from Moscow Engineering Physics Institute,
     Moscow, 115409,  Russia}$,
G.~Wang$^4$, J.~P.~Wessels$^6$, C.~L.~Woody$^1$, N.~Xu$^6$,
Y.~Zhang$^6$, C.~Zou$^6$\\ (E877 Collaboration)}
\address{$^1$ Brookhaven National Laboratory, Upton, NY 11973\\
$^2$ Gesellschaft f\"ur Schwerionenforschung, Darmstadt, Germany\\
$^3$ Idaho National Engineering Laboratory, Idaho Falls, ID 83402\\
$^4$ McGill University, Montreal, Canada\\
$^5$ University of Pittsburgh, Pittsburgh, PA 15260\\
$^6$ SUNY, Stony Brook, NY 11794\\
$^7$ University of S\~ao Paulo, Brazil\\
$^8$ Wayne State University, Detroit, MI 48202\\}
\date{\today}
\maketitle

\begin{abstract}
Experimental results and a detailed analysis are presented of the
transverse energy and charged particle azimuthal distributions measured
by the E877 collaboration for different centralities of Au+Au collisions
at a beam momentum of 10.8 A GeV/c. The anisotropy of these
distributions is studied with respect to the reaction plane
reconstructed on an event-by-event basis using the transverse energy
distribution measured by calorimeters.  Results are corrected for
the reaction plane resolution. For semicentral events we observe
directed flow signals of up to ten percent. We observe a stronger
anisotropy for slow charged particles. For both the charged particle and
transverse energy distributions we observe a small but non zero elliptic
anisotropy with the major axis pointing into the reaction
plane. Combining the information on transverse energy and charged
particle flow we obtain information on the flow of nucleons and
pions. The data are compared to event generators and the need to
introduce a mean field or nucleon-nucleon potential is discussed.
\end{abstract}
\narrowtext

\section{Introduction}
Collisions between two gold nuclei of about 11 A$\cdot$~GeV/c momentum
at the AGS 
have been characterized rather completely in terms of the global
observables, transverse energy ${\rm E_T}$ \cite{877et} and charged
particle multiplicity ${\rm N_c}$ \cite{877nc}. The picture that emerged
from these measurements is that the two gold nuclei stop each other to a
very high degree. Through comparison to models that reproduce the
experimental observables initial particle and energy densities have been
inferred and maximum values around ten times normal nuclear matter
density and 2 GeV/fm$^3$ have been found \cite{rqmd,arc,kahana}.  On the other
hand, hadrons cease to interact strongly and freeze-out at a density
significantly below nuclear matter density (for Si + Au collisions at
the AGS see \cite{thermal}). The interesting question arises to what
degree the system loses its memory of the initial highly compressed
phase during the subsequent expansion.

While particle yields are consistent with chemical equilibrium already
for the lighter Si~+~Au system \cite{thermal}, particle spectra show that the
equilibrium is only local and that overall the system expands
longitudinally and transversely with an average velocity of one half and
one third of the speed of light \cite{thermal}, respectively. Recently we
found from analyzing the azimuthal asymmetry of the transverse energy
distribution that the system even remembers the initial collisions
geometry or the impact parameter: for all but the most peripheral and
the most central collisions a dipole component also called `sideward
flow' is observed in the transverse energy azimuthal distribution
forward and backward of mid-pseudorapidity \cite{flow}. The forward and 
backward flow effects are back-to-back or 180$^o$ relative to each
other. The effect is largest in semicentral collisions. Integrating over
pseudorapidities forward of $\eta$ = 1.85, about 9\% of the transverse
energy is carried by this directed flow \cite{flow}.

Following up on the initial discovery of this sideward flow at AGS
energies our goal is to characterize the effect in more detail in order
to eventually gain access to the equation of state of nuclear matter at
the extreme densities reached initially in gold-gold collisions at 11
A~$\cdot$~GeV/c. In this paper, we present a complete characterization of
the flow behavior in transverse energy and charged particle multiplicity
with fine binning in pseudorapidity and as a function of centrality of
the collision. At the same time we are studying the triple differential
cross section of the emission of identified particles such as protons and
pions in the E877 forward spectrometer \cite{flow-qm95}. This will be
the subject of a future publication. In the following
Chapter we will briefly describe the experimental set-up and conditions,
and introduce the analysis method in Chapter 3. The resulting
anisotropies are presented in Chapter 4. In Chapter 5, we analyse the
flow signal from one of the cascade codes (RQMD \cite{rqmd-gen}) that
describe the collisions in terms of individual hadron-hadron collisions
and compare this prediction to the experimental data. In Chapter 6 we
use the complementarity of the two measurements (in multiplicity ${\rm
N_c}$ and in transverse energy ${\rm E_T}$) to disentangle the
contribution of pions and nucleons to the observed flow signals and to
again compare to model predictions as well as to lower energy data.  

\section{Experiment}

In the experiment presented here a 10.8~A$\cdot$~GeV/c Au beam of the
Brookhaven AGS was impinging on Au targets of 540 and 980 mg/cm$^2$
areal density corresponding to 1.07 and 1.94~\% of a Au + Au nuclear
interaction length. The reaction products were detected in the E877
apparatus schematically depicted in Figure~1. The detectors used in the
present analysis are shown enlarged in the insert. In the fall 1993, AGS
heavy-ion run information from about 10$^7$ Au + Au collisions was
collected sampling the whole impact parameter range with parallel
triggers requiring different levels of ${\rm E_T}$ or just the presence
of a beam particle.

Every incident beam particle is characterized by the scintillator
hodoscope (S1-S4), and the horizontal position and angle of incidence at
the target are measured by a pair of silicon microstrip detectors
(BVER1,2). The method used to correct for the beam displacement and
direction is described in detail in \cite{877nc}. The angular
divergence of the beam is of the order of 1 mr and much smaller than the
bin width in $\eta$ and $\phi$ used in the present
analysis. Interactions occurring upstream of the target are effectively
rejected by requiring (i) that the pulseheight measured in a 100 $\mu$m
thick silicon detector just upstream of the target is consistent with
the energy loss of a Au ion, (ii) that a (beam) particle in BVER1,2 is
not accompanied by other tracks, and (iii) that the correlation between
${\rm E_T}$ measured in different ranges of pseudorapidity $\eta$
follows the systematics for interactions in the target (see below).
 
The event characterization is obtained using the transverse energy
measured in the two calorimeters surrounding the target, the target
calorimeter (TCal) and participant calorimeter (PCal). The TCal consists
of NaI crystals of 5.3 radiation lengths depth. In the present analysis
the pseudorapidity range -0.5 $\le \eta \le 0.8$ is used to measure
transverse energy in 13 $\times$ 64 bins in $\eta$ and azimuthal angle
$\phi$. For more details on the TCal and the analysis of TCal data see
\cite{877et,814et,814fb}. The ${\rm E_T}$ measurement at central and
forward rapidity is obtained using the PCal, a lead/iron/scintillator
sampling calorimeter described in \cite{877et,pcal}. The PCal has full
azimuthal coverage with a granularity of $\Delta \phi = 20^{\circ}$. In
pseudorapidity data are obtained for 17 bins covering 0.9 $\le\eta \le$
4.2 (see Figure~2). The 4 
depths sections of the calorimeter are not used separately in the
present analysis. The orientation of the reaction plane is determined
event-by-event using ${\rm E_T}$ from the TCal or one of several $\eta$
regions of the PCal. The azimuthal distribution of ${\rm E_T}$ relative
to the reaction plane is then determined over the full range of -0.5
$\le \eta \le $ 4.2.

The distribution of charged particles ${\rm N_c}$ is measured in two
identical silicon pad detectors of 300 $\mu$m thickness. The placement
and segmentation of the silicon detectors is shown in Figure~3. Each
detector is segmented into 512 pads with 8 $\times$ 64 bins in $\eta$
and $\phi$ of typically 100 mr width. The azimuthal distribution of the
charged particle emission relative to the reaction plane is measured in
12 bins for 0.8 $\le \eta \le $ 2.65. An analysis of the ${\rm N_c}$
distribution as a function of centrality in Au + Au collisions is
published in \cite{877nc}. Details of the analysis technique, e.g. how
to deal with $\delta$-rays, multiple hits, beam displacement, can be
found there and were adopted for the present analysis as well. The
present data are not corrected for $\gamma$ conversion (carrying the
flow information of $\pi^0$) which in the analysis presented below
accounts for about 6 \% of the hits in the silicon pad detectors.

\section {Flow Analysis}
The azimuthal anisotropy is analysed as a function of the centrality of
the collision. Centrality is measured by ${\rm E_T}$ in the
calorimeters. Figure~4a displays for both detectors the fraction of the
geometric cross section $\sigma_{top}({\rm E_T})/\sigma_{geo}$ obtained by
integrating from a given value of ${\rm E_T}$ to the maximum ${\rm E_T}$
observed. Here the geometric cross section is defined as ${\rm
\sigma_{geo} = \pi r_0^2(A^{1/3}+A^{1/3})^2 = 6.127}$ b with the mass
number A = 197 and ${\rm r_0}$ = 1.2 fm. Both distributions are not
unfolded for detector response. The shape of the distributions
is very similar for the two detectors except that the fall-off for very
central collisions is somewhat wider for the TCal because of the smaller
$\eta$ coverage and larger leakage fluctuations. Figure~4b shows the
projection of the correlation between the two ${\rm E_T}$ measurements
on either axis with error bars indicating the width (standard deviation)
of the correlation. The correlation is close to linear over most of the
range and only for collisions in the top 5~\% range of centrality does
one or the other centrality measure select different events. Also shown
in the figure are the ${\rm E_T}$ bins used in the analysis. The most
peripheral bins start at 5 and 50 GeV in TCal and PCal, respectively,
i.e. only the top half of the geometric cross section is studied. The
results presented later have been corrected for interactions not
occuring in the target and the correction to the resulting anisotropies
is noticable only for the more peripheral collisions with PCal ${\rm E_T
\le}$ 100 GeV.

In our previous analysis of the azimuthal anisotropy of ${\rm E_T}$
production \cite{flow} we have subdivided the data into $\eta$ bins and have
performed, event-by-event, a Fourier analysis \cite{flow-zv} of the
azimuthal distribution in each $\eta$ bin. This method has the advantage
that it involves only one $\eta$ interval at a time and that it does not
require to determine a reaction plane angle. Hence it is not influenced
by the resolution with which different detectors can measure the
reaction plane angle. However, since the Fourier analysis is performed
for every event, the size of the $\eta$ bin has to be large enough to
allow to distinguish a true anisotropy from a statistical
fluctuation. In central Au + Au collisions the total multiplicity
reaches indeed large values of 800 - 900 over the full solid
angle. However, first results on the centrality dependence of the
anisotropy \cite{flow} found the effect to be maximal in semi-central
collisions where the multiplicity is significantly lower. In practice,
this was limiting our previous analysis to three $\eta$ bins. Since we
did indeed see a pronounced first (or dipole) moment with a strong
back-to-back correlation of forward and backward $\eta$ bins we choose a
different strategy in the present analysis to now study the flow effects
in small $\eta$ bins.

From the data, the azimuthal angle $\Psi_n^{(i)}$ of the n-th moment of the
transverse energy distribution in the $\eta$ window $i$ is obtained via
\begin{equation}
\tan \Psi_n^{(i)} = \frac
{{\displaystyle \sum_{j}}(\pm){\rm E_T}^j \sin n\phi_j}
{{\displaystyle \sum_{j}}(\pm){\rm E_T}^j \cos n\phi_j} = \frac
{{\displaystyle \sum_{j}}(\pm){\rm E_{Tx}}^j}
{{\displaystyle \sum_{j}}(\pm){\rm E_{Ty}}^j}
\end{equation}
where the sum runs over the $j$ cells with azimuthal angles $\phi_j$ of
the detector in an $\eta$ window $i$ and the sign is positive (negative) for
cells at $\eta$ forward (backward) of mid-pseudorapidity. For n=1 this
is the equivalent of the directivity method used in \cite{fopi}, except
that we use ${\rm E_T}$ instead of ${\rm p_t}$.

For every event, the angle $\Psi_1^{(i)}$ of the dipole component is
found in the i-th of four pseudorapidity windows. The most backward
window W1 covers the range -0.5 $\le \eta \le$ 0.7, where the TCal has
full azimuthal coverage. The windows W2, W3, and W4 label regions of the
PCal covering approximately 0.8 $\le \eta <$ 1.4, 2.0 $\le \eta <$
2.7, and 2.7 $\le \eta <$ 4.5. Here the region around midrapidity is
intentionally skipped since the dipole component is expected to cross
zero in that region. We denote by $\Psi_R$ the angle the reaction plane
(defined by the impact parameter ${\rm \vec{b}}$ and the beam direction
\^z) makes with the laboratory x-axis (see Figure 1). The angles
$\Psi_1^{(i)}$ are 
the experimental measure of $\Psi_R$. A remaining twofold ambiguity is
solved by defining that in the forward hemisphere $\Psi_1$ points in the
direction of ${\rm \vec{b}}$, where ${\rm \vec{b}}$ points from target
to projectile. This is consistent with the assumption that the
projectile scatters away from the target (repulsive trajectory).
The angle
$\Psi_1$ is shifting by a phase of $\pi$ at midrapidity. We have
experimentally verified this back-to-back correlation
\cite{flow}. Neglecting the phase information, $\Psi_1$ is generally called
the reaction plane angle and we will stay with this terminology. Since
there is only one reaction plane orientation for every collision a
comparison of the angles $\Psi_1^{(i)}$ measured in the four $\eta$
windows allows to extract the resolution with which $\Psi_R$ is measured
in each window.

The azimuthal distribution with respect to the reaction plane angle
of a global observable $X$ is expanded in terms of its Fourier
components  
\begin{equation}
\frac{{\rm d}^2X}{{\rm d\eta d}(\phi-\Psi_R)}
 = v_0 (1 + \sum_{n \geq 1} (2 v_n \cos n(\phi-\Psi_R)),
\end{equation}
where $v_0 = \langle X \rangle_{\eta}/2\pi$ and $\langle X \rangle_{\eta}$ is
the average of the observable $X$ in the pseudorapitiy interval
$d\eta$. Note that sine terms are missing because of the necessary
reflection symmetry with respect to the reaction plane. This expansion
is equivalent to a decomposition into multipole components in a plane
(transverse to the beam direction).

A Fourier decomposition of the distribution $X$ measured with respect to
the reaction plane angle determined in the $i$-th window yields
\begin{equation}
\frac{{\rm d}^2X}{{\rm d\eta d}(\phi-\Psi_1^{(i)})}
 = v_0 (1 + \sum_{n \geq 1} (2 v_n' \cos n(\phi-\Psi_1^{(i)}))
\end{equation}
and for practical reasons we limit the analysis to $n$=1,2 (see below).
The Fourier coefficients in this series are evaluated by fitting
equation (3) to the data or from:
\begin{equation}
v_n'=\frac{
\langle {\displaystyle \sum_{k}}X^k  \cos n(\phi_k-\Psi_1^{(i)}) \rangle}
 { \langle {\displaystyle \sum_{k}}X^k \rangle},
\end{equation}
where the sum is taken over all cells of the detector belonging to a
pseudorapidity bin under study, and the brackets refer to the event
average evaluated for a given event class (centrality).
 
From the measured Fourier coefficients $v_n'$ the true values $v_n$ can
be obtained (see also~\cite{flow-zv}) by unfolding for the finite
resolution with which $\Psi_R$ is measured, using
\begin{equation}
v_n = \frac{v_n'}{|\langle \cos n(\Psi_1^{(i)}-\Psi_R) \rangle|}.
\end{equation} 
Again the brackets indicate the event average evaluated for a given
pseudorapidity window and a given event class (centrality). The
measurement of $\Psi_1^{(i)}$ in three or more pseudorapidity windows
(four in  our case) allows to evaluate the correction factors $\langle
\cos n(\Psi_1^{(i)}-\Psi_R) \rangle $ directly from the data without
further assumptions. We have, e.g. for n=1,  
\begin{equation}
\cos (\Psi_1^{(i)}-\Psi_1^{(j)}) = \cos (\Psi_1^{(i)}-\Psi_R) \cos
(\Psi_1^{(j)}-\Psi_R) + \sin (\Psi_1^{(i)}-\Psi_R) \sin
(\Psi_1^{(j)}-\Psi_R).
\end{equation}
Taking the event average, using the reflection symmetry of the $\phi$
distribution with respect to the
reaction plane and assuming that the only correlation between
pseudorapidity windows $i$ and $j$ is via the flow effect we obtain
\begin{equation}
\langle \cos (\Psi_1^{(i)}-\Psi_1^{(j)}) \rangle = \langle \cos
(\Psi_1^{(i)}-\Psi_R)\rangle \langle \cos (\Psi_1^{(j)}-\Psi_R)\rangle.
\end{equation}
Combining these equalities for the four pseudorapidity windows we can 
evaluate the effect of the finite reaction plane resolution in window
$i$ as a function of centrality. The results are shown in Figure~5. The
resolution in a given $\eta$ interval is determined by the finite
granularity, the energy resolution and leakage fluctuation of the
detector, and the magnitude of the anisotropy in this $\eta$ interval.
The symbols in the left diagram reflect the correction to be applied to
the measured dipole component. The correction is smallest for
semicentral collisions (PCal ${\rm E_T \approx}$ 220 GeV) where we found
the flow effect to be largest \cite{flow}. Comparing the different
pseudorapidity windows, the resolution is best for the most forward
window W4 but W3 and W1 also give satisfactory results. In the window W2
the correction is rather sizeable as expected, more than a factor of two
for all centralities, and we discard for the following analysis this
window for purposes of reaction plane determination. The signs of the
correction factors reflect the phase shift by $\pi$ in the angles
$\Psi_1$ at mid-pseudorapidity.

The effect of finite reaction plane resolution becomes more significant
for higher multipole components as displayed on the right hand side of
Figure~5 for the quadrupole component. There, only the two forward PCal
windows yield manageable corrections of about a factor two for
semicentral collisions and a factor four to six for central
collisions. This shows the difficulty to extract any multipole
components with $n \geq$ 3 from the data by methods involving the
determination of a reaction plane. Our previous method \cite{flow} of
event-by-event Fourier decomposition does not have this limitation but
is limited by the finite multiplicity in an event which depends {\it
e.g.} on the beam energy and centrality of the collision.

One may ask to what extent the accuracy of the correction is affected by
remaining detector imperfections such as, e.g., miscalibrated or missing
calorimeter channels which may bias the distribution of
$(\Psi_1^{(i)}-\Psi_1^{(j)})$. We have studied this question by
generating a probability distribution in angle difference normalized to
a probability distribution from 'mixed events' where the two angles are
from different events. Using this probability distribution in the
averaging procedure yields, for W4 for instance, the histogram presented
in Fig. 5 as compared
to the points. The differences are small, typically 5~\% or less for all
four pseudorapidity windows.

Analyzing the calorimeter data care has to be taken to assign the proper
pseudorapidity value to each tower of the calorimeter.  The spread of
showers, the nonprojective geometry of a detector, and the variation in
the ${\rm E_T}$ distribution over the solid angle covered by a detector cell
will, in general, result in an effective mean pseudorapidity, which is
different from the pseudorapidity of the center of the tower.  As in
\cite{877et} we have simulated the PCal performance using the
GEANT~\cite{geant} package combined with an event generator that
reproduces the measured ${\rm E_T}$ distribution and with a fast shower
deposition code PROPHET~\cite{prophet}.  The pseudorapidity
distributions of the particles which contribute to each PCal tower were
calculated.  The mean value of pseudorapidity weighted with the
deposited energy was determined and used later in the analysis as the
tower pseudorapidity.  In Figure~6 we show how the assigned $\eta$ values
differ from the pseudorapidity of the cell geometrical center.  We also
show in this figure the spread of $\eta$ values (standard deviation) of
particles contributing to the energy deposit in a given tower. This
gives an indication that structures in the azimuthal anisotropy of the
${\rm E_T}$ distribution cannot be resolved to better than about 0.5
units of pseudorapidity.  Using different, realistic event generators
and different centralities of the collision we checked that the assigned
pseudorapidity values are not visibly model or centrality
dependent. The differences in mean values for different event
generators/centralities are much less than the widths shown in the
figure, and for the midrapidity region were found to be less than 0.05.

\section{AZIMUTHAL DISTRIBUTIONS OF CHARGED PARTICLES AND TRANSVERSE ENERGY}

\subsection{Charged particles}

The azimuthal distribution of the charged particle multiplicity is
studied for five bins in centrality and with a reaction plane
orientation determined using the TCal and the most forward PCal section to
avoid auto-correlations. As an example of such a double differential
distribution the data for the intermediate centrality bin are shown in
Figure~7 both in a three-dimensional representation and as a few slices
at certain $\eta$ values. A pronounced dipole component and its sign
change at mid-pseudorapidity are immediately obvious. Closer inspection
reveals in addition a quadrupole component, easily visible {\it e.g.} in
Figure~7 around $\eta$ = 1.7 where the dipole moment vanishes. Figure~8a
shows the corrected first and second moments of the Fourier
decomposition for all five centrality bins. The error bars reflect for
each centrality the typical statistical errors as well as systematic
errors connected to variations of the experimental conditions during the
run (e.g. beam position). These were obtained by
subdividing the entire data sample into subsamples (runs of 100 k events)
and obtaining the standard deviation of the results from these
subsamples. The two different windows used to determine the reaction
plane lead, after correction for resolution, to very similar results. From
this comparison we conclude that the relative systematic errors
in the corrected coefficients $v_1$ and $v_2$ which are mostly
determined by the correction for the reaction plane resolution are less
than 10 and 20~\% , respectively. For very small values of $v_1$ and
$v_2$ we estimate absolute systematic errors of 0.005.

The finite dipole component $v_1$ represents directed sideward flow of
charged particles in qualitatively the same way as seen in our previous
study of ${\rm E_T}$ \cite{flow}. The dipole component shows a
characteristic zero-crossing around mid-pseudorapidity and is nonzero
elsewhere for all centralities chosen. The sign of the charged particle
flow is such that on average charged particles go in the same direction
as the transverse energy. However, the anisotropy is small, at most
0.03. There is a subtle change in shape of the $\eta$ dependence and in
the location of the zero crossing with centrality.  

We also find a nonzero quadrupole component which is even smaller, at
most 2~\% after correction. But the deviation from zero is significant
as can be judged from the projection in Figure~7. There is no
visible pseudorapidity dependence of $v_2$. The positive values of
$v_2$ imply enhanced yields in the reaction plane. Hence, the small
quadrupole component we find is oriented perpendicular to the 'squeeze-out'
observed at lower beam energies in the 1-2~A$\cdot$~GeV/c range
\cite{squeeze_1} where preferential emission out of the reaction plane
was established.

In a further analysis step, we determine the anisotropy of only those
tracks that deposit more than four times the minimum ionizing energy
loss in the silicon pad detector. This selects mostly low momentum
particles, preferentially slow protons. The resulting anisotropy
parameters are displayed in Figure~8 (right panel). Two general trends
are noticeable 
as compared to the results for all charged particles displayed in
Figure~8 (left). i) The magnitude of the anisotropy is significantly bigger,
reaching values up to 10~\%. ii) The location of the zero-crossing
shifts forward in pseudorapidity. This is expected because of the
difference between rapidity and pseudorapidity for more massive (less
relativistic) particles combined with the fact that protons dominate
this data sample while they account overall for roughly 1/3 of all
charged particles. The larger magnitude of the anisotropy for more
heavily ionizing particles could indicate
that protons exhibit a stronger sideward flow effect than pions.
 
\subsection{Transverse energy}

A similar event shape analysis was performed on the transverse energy
combining data from the two calorimeters TCal and PCal thus covering the
range -0.5 $\leq \eta \leq$ 4.2. We fit the experimental $\phi$
distribution of ${\rm E_T}$ relative to a reaction plane determined with
any of the three windows W1, W3, and W4 that do not overlap in
pseudorapidity with the ${\rm E_T}$ bin with the functional form
\begin{equation}
\frac{{\rm d^2 E_T}}{{\rm d\eta d}(\phi-\Psi_1^{(1,3,4)})}
 = v_0 (1 + \sum_1^2 (2 v_n' \cos n(\phi-\Psi_1^{(1,3,4)})).
\end{equation}
This is done for 15 centrality bins gating on PCal ${\rm E_T}$ ranges as
indicated in Figure~4. After unfolding the coefficients $v_n'$ for the
reaction plane resolution two or three values are available for every
$\eta$ from the reaction plane measurements not overlapping in $\eta$.
This provides a good check on the systematics. In order to correct for
any asymmetries caused by interactions other than in the target we also
evaluate the anisotropy coefficients from special {\it target-out}
runs. The correction matters only for the two most peripheral
centrality bins. In the first (second) bin it is found that the absolute
correction to $v_1$ is of the order of 0.005 (0.001). As in the case
of the charged particle analysis the 
systematic error is dominated by the accuracy of the correction for the
reaction plane resolution and we assign a 10~\% relative systematic error or an
absolute systematic error of 0.005 to the
corrected dipole coefficients.

Figure~9 shows the resulting dipole coefficients for a representative
sample of centralities. Statistical errors were obtained in the same way
as in the charged particle multiplicity analysis. The data were divided
into 12 subsamples and the scatter of results from these subsamples
defines the error of the mean. As a function of pseudorapidity the data
in the 24 experimental bins form a quasi-continuous distribution with a
smooth evolution from negative to positive values for more forward
$\eta$ with a zero crossing around $\eta$ = 1.9. The data shown in
Figure~9 can be compared to the values for three large $\eta$ windows
(-0.5 $\leq \eta \leq$ 0.8, 0.83 $\leq \eta \leq$ 1.85, and 1.85 $\leq
\eta \leq$ 4.7) used in 
our first analysis \cite{flow}. With the much finer segmentation in
$\eta$ it is now possible to verify that the vanishing of $v_1$ in the
middle $\eta$ window is indeed due to the zero crossing of $v_1$ around
mid-pseudorapidity as we had suspected.

The evolution of $v_1$ as a function of centrality shows several
systematic features. The location of the zero-crossing at $\eta$ = 1.9 -
2.0 does not depend significantly on centrality except for the two most
peripheral bins where more forward values are observed. The dependence
of $v_1$ on pseudorapidity is characterized by an s-shaped curve with a
minimum around $\eta \approx 0$ and a maximum around $\eta \approx
3.0$. Inspecting these extrema in $v_1$ as function of centrality, they
reach maximum values in the range 130 - 270~GeV corresponding to collisions
in the top 30 - 5~\% centrality region. The maximum flow values backward
and forward are 7 and 12~\% respectively. Furthermore, the shape of the
distributions in Figure~9 changes with centrality; the extrema of the
s-shaped curve move closer to midrapidity for increasing centrality.

At lower beam momenta the slope of $v_1$ at midrapidity has been used to
quantify the strength of the flow effect. For the present data the value
is d$v_1$/d$\eta$ = d(${\rm \langle E_x \rangle/\langle E_T \rangle
}$)/d$\eta \approx$ 0.07 around 15~\% centrality where the flow effect
is maximal, and it decreases to 0.04 for the highest centrality bin
studied here. The values of the slope are significantly smaller than
reported at lower energies \cite{low_flow} for a similar quantity,
d(${\rm \langle p_x \rangle/\langle p_t \rangle }$)/dy, evaluated for
protons. At beam kinetic energies per nucleon of 150, 250 and 400 MeV
values of d(${\rm \langle p_x \rangle/\langle p_t \rangle }$)/dy = 1.43,
1.23 and 1.22 have been obtained. The increase in ${\rm 
\langle p_t \rangle }$ or ${\rm \langle E_T \rangle }$ and the increase
in y make it plausible that the relative strength of the flow is smaller
at AGS energies. Below we discuss a procedure to separate the flow effect of
pions and nucleons and to relate $\eta$ and y to obtain a more
quantitative understanding of the systematics of the observed strong
energy dependence. The comparison to lower energy data is resumed there.

Figure~9 together with the d${\rm E_T/d\eta}$ distribution \cite{877et}
shows where 
the most sensitive $\eta$ intervals are to determine the orientation of the
reaction plane: $\eta \lesssim 0.8$ and $\eta \gtrsim 3.0$. This is in line
with the results shown in Figure~5 for the experimental reaction plane
resolution.

Figure~10 presents the results for the quadrupole component of the ${\rm
E_T}$ distribution. For intermediate centralities (130-270 GeV,
corresponding to the top 5-30~\% of the geometric cross section) small but
significant values of 1-2~\% are observed. For more central and more
peripheral collisions they decrease to zero. The statistical errors are
shown in the figure, the relative and absolute systematic errors are
estimated to be 20~\% and 0.5~\%. Quantitatively the values are very
similar to the quadrupole  
anisotropy observed in the charged particle distributions. Again there
is no significant dependence on pseudorapidity and again the values are
positive, i.e. emission is enhanced in the reaction plane, not
perpendicular to it.

\section{COMPARISON WITH MODELS}

Both global distributions in ${\rm E_T}$ and ${\rm N_c}$ and spectra of
identified protons and pions have been compared
\cite{877et,877nc,877rao,866shig,877prpi} to predictions from two event
generators based on hadronic cascades, RQMD \cite{rqmd} and ARC
\cite{arc,kahana}. Although some discrepancies are noted, in particular a
peaking in d${\rm E_T}$/d$\eta$ too much forward as compared to the data
and proton spectra significantly steeper than the data close to
midrapidity (for ARC only a spectrum half a unit away from midrapidity
has been published \cite{kahana}), the overall agreement otherwise is
good. The slope of the proton spectra can be linked to transverse
expansion of the system \cite{betaside}. Analysis of the RQMD freeze-out
condition indicates \cite{rqmd_bt,mattiello} that, in the cascading of many
successive hadronic collisions, a collective transverse expansion is
built up, but apparently for Au + Au collisions at AGS energies the model
in its cascade mode underpredicts the transverse expansion
velocity. With the present data 
we can subject the models to a different test of the collective
velocities at freeze-out. By evaluating the sideward flow in the same
manner as in the present analysis and comparing to the data we
test the anisotropic component of the expansion, i.e. the component that
carries the memory of the impact parameter and therefore may be
sensitive to the equation of state of the system.

Figure~11 shows the dipole component of the azimuthal distribution of
transverse energy and charged particle multiplicity for collisions of
intermediate centrality (${\rm \sigma_{top}/\sigma_{geo}}$ = 5-15\%)
both from experiment and evaluated from events simulated with RQMD.  It
is apparent that the experimental anisotropies of both the charged particle
and of the transverse energy distributions are quantitatively quite
different with values for ${\rm E_T}$ about twice the values for ${\rm
N_c}$. The extracted flow
parameters $v_1$ for a given particle species may 
differ depending on whether they were extracted from azimuthal
distribution of the number density or the transverse energy density.
It will be shown in the next section that this difference is 
not the dominant part of the effect seen. Since pions and nucleons
contribute with different relative 
weights to ${\rm E_T}$ (composed mainly of energy deposits of p,n,$\pi^+,
\pi^-, \pi^0$) and to ${\rm N_c}$ (counting essentially the number of p, 
$\pi^+, \pi^-$) one suspects that the observed difference in anisotropy
is due to a different behavior of pions and nucleons. Another indication for a
difference between pions and nucleons is the different dipole anisotropy
seen for all charged particles and heavily ionizing particles (see
discussion above and Fig.~8).

RQMD reproduces neither the experimental anisotropy for ${\rm E_T}$ nor for
${\rm N_c}$ but, in agreement with the data, there is a difference
between the 
two with, in general, more positive values of $v_1$ for the ${\rm E_T}$
distribution. In the model we can separate the contribution from pions
and nucleons and Figure~12 shows, for the same centrality range as in
Figure~11, the dipole anisotropy for protons, pions and, for comparison,
also the anisotropy of ${\rm E_T}$ and ${\rm
N_c}$. Protons and pions show opposite flow effects of the
same order of magnitude leading to differences and even to a change in
sign between the dipole anisotropy of ${\rm E_T}$ and ${\rm N_c}$ in the
range $\eta$ = 2 - 3. Qualitatively this is in agreement with the
feature exhibited by the data but quantitatively the model does not
reproduce the data. The failure of RQMD to account for the anisotropy in
${\rm E_T}$ was already apparent in our first analysis in three large
$\eta$ bins \cite{flow}. A comparison of data and model for the bin
$\eta$ = 1.85 - 
4.7 showed \cite{flow-qm95} that the model underpredicts the
experimental dipole component by a factor of two. This discrepancy
combined with the possible intricate cancellations of flow effects of
pions and nucleons, as shown in Fig.~12, provides another
motivation to separate the experimental effect according to particle
species.

We have also evaluated from the RQMD simulations the quadrupole
anisotropy coefficients $v_2$ for semicentral collisions. They are found
to be very close to our experimental observation, with typical positive
values of 0.01 - 0.02, and no significant pseudorapidity
dependence. Further, we find that both pions and nucleons are
contributing to this anisotropy with equal sign and comparable
magnitude.

As far as the ARC event generator is concerned, a more limited
comparison with the present data is possible by inspecting results of
calculations shown in a recent preprint \cite{arc_flow}. In this work, a
comparison is made to the results of our first analysis in three coarse
pseudorapidity bins and it appears that for the forward bin the dipole
component is rather well reproduced by ARC. It is interesting to note
that this agreement is achieved by introducing an energy dependent
treatment of the nucleon-nucleon scattering with a gradual transition
from repulsive scattering at low relative energies to an equal
probability for repulsive and attractive trajectories at higher
energies. Using only the latter without energy dependence the flow for
protons is reduced to half it value. Another interesting feature emerges
from the ARC simulations. There the protons exhibit a quadrupole
anisotropy with the long axis perpendicular to the reaction plane and
this anisotropy vanishes as beam rapidity is approached. Both features
are at variance with the present experimental observation of the
orientation (in-plane) and pseudorapidity independence of $v_2$.
 
\section{FLOW OF NUCLEONS AND PIONS}

Using the present data on flow of transverse energy and charged
particles we can try to separate the contribution of nucleons and pions
to the flow effect. In this analysis we assume that the observed flow in 
the global observables ${\rm 
E_T}$ and ${\rm N_c}$ is a linear superposition of the anisotropy of
nucleons and pions, thereby neglecting the contribution from other particle
species. We denote the respective flow parameters by $v_1^{({\rm N_c})}$ and
$v_1^{{\rm (E_T)}}$ and further differentiate between coefficients 
$v_1^{({\rm N_c,n})}$ $v_1^{({\rm N_c},\pi)}$, $v_1^{({\rm E_T,n})}$, and
$v_1^{({\rm E_T},\pi)}$ for nucleons and pions, respectively.
The dipole anisotropy of the two global observables can then be
written as:

\medskip

\begin{eqnarray}
v_1^{({\rm N_c})}&=& \frac{{\rm dN_c^{\pi}/d}\eta \; \cdot v_1^{({\rm
N_c},\pi)} + {\rm dN_c^{n}/d}\eta \; \cdot v_1^{({\rm N_c,n})}} 
          {{\rm dN_c^{\pi}/d}\eta + {\rm dN_c^{n}/d}\eta },
\\
v_1^{({\rm E_T})}&=& \frac{{\rm dE_T^{\pi}/d}\eta \; \cdot v_1^{({\rm
E_T},\pi)} + {\rm dE_T^{n}/d}\eta \; \cdot v_1^{({\rm E_T,n})}} 
           {{\rm dE_T^{\pi}/d}\eta + {\rm dE_T^{n}/d}\eta }.
\end{eqnarray}
These equations can be solved for the flow parameters of pions and
nucleons if one knows in addition to the measured $v_1^{({\rm N_c})}$ and
$v_1^{({\rm E_T})}$ values: 
(i) the relative contribution of pions and nucleons to the charged particle
and transverse energy pseudorapidity distributions, and
(ii) the ratio of the flow parameters for a given particle species
arising from particle or 
${\rm E_T}$ azimuthal distributions, i.e. $v_1^{(N_{c})}/v_1^{({\rm
E_T})}$ for pions and nucleons separately. 

Proton and pion spectra have been measured for the top 7 \% of
centrality over nearly 4$\pi$ if one employs symmetry
with respect to midrapidity and combines data from E866
\cite{866shig,866aki} and E877 \cite{877prpi}. We have parameterized the
measured rapidity distributions of protons and pions as Gaussian distributions
and the transverse mass distributions as Boltzmann
distributions with slope constants that again have a Gaussian
distribution as a function of rapidity. This provides the double
differential cross section $d^2 \sigma/{\rm dy dm_t}$ for protons and pions
for the full phase space. From this information distributions of transverse
energy or charged particle multiplicity can be computed for any
pseudorapidity. In order to test the quality of the parameterization and
also the internal consistency of the data we have compared the
distributions ${\rm dE_T/d}\eta$ and ${\rm dN_c/d}\eta$ from the
parameterization 
to the quantities measured by E877 \cite{877et,877nc} with an entirely
different detector system for the same centrality and excellent
agreement is found for both quantities.  As an alternative check we have
used the relative contribution of nucleons and pions to ${\rm E_T}$ and
${\rm N_c}$ from RQMD and the absolute difference in the resulting
values of $v_1$ for nucleons and pions is less than 0.005.

To estimate the difference in the anisotropy coefficient measured for
${\rm E_T}$ and ${\rm N_c}$ we again have used two approaches. We assume
that the anisotropy (flow) is due to a displacement of the triple
differential cross section ${\rm d^3 \sigma/dp_x dp_y dy}
$ by some
rapidity dependent amount ${\rm p_{x0}(y)}$. This is close to our present
experimental observation \cite{flow-qm95}. For moderate
displacements (${\rm p_{x0}} \leq$ 0.15 GeV/c; well justified in the
rapidity range and for the system considered here) and a Gaussian
distribution in ${\rm p_x, p_y}$ one can show that the
ratio between $v_1^{{\rm E_T}}$ and $v_1^{{\rm N_c}}$ is
4/$\pi$. To check the influence of this assumption on the resulting pion
and nucleon flow, we have used events from RQMD to numerically
evaluate this quantity. The resulting values $v_1$ for nucleons and
pions are smaller by typically 0.01 (absolute difference).

With the two ingredients i) and ii) such determined and the measured
flow parameters (left hand 
side of equations (9) and (10)) we can solve equations (9) and (10) for
every pseudorapidity to extract $v_1^{{\rm E_T,n}}$ and $v_1^{{\rm
E_T,\pi}}$. The resulting flow parameters for nucleons and
pions are shown in Figure~13. One can see that indeed the difference in
the flow parameters of transverse energy and charged particle
multiplicity can be attributed to a distinctly different behavior of
nucleons and 
pions. Nucleons show a pronounced flow effect, pions show a much weaker
effect and a tendency to preferentially be emitted to the side opposite of the
protons. The assumptions made in this analysis lead to a correlated
systematic error in the resulting flow coefficients for
nucleons and pions and we estimate relative errors of 10 and 50 \%,
respectively. The uncertainty for 
pions becomes relatively large because the anisotropy found is so small.

A comparison of the data for nucleons and pions to the corresponding
quantities from RQMD is also given in Fig.~13. This allows to understand
the discrepancy between data and model for the global observables. The proton
flow is underpredicted by the model at forward pseudorapidities and at
the same time a stronger trend for pions to go the opposite way is
predicted. This latter feature has been dubbed ``antiflow'' in the
literature, a somewhat misleading term since the effect (in the code) is due to
shadowing. The combination of these two effects (underprediction of
nucleon and overpredictin of opposite pion flow) leads to a transverse
energy flow close to zero in the model for pseudorapidities less than 3
while the data show a pronounced flow effect there. At backward
pseudorapidities in the model proton flow and pion shadowing nearly
cancel. In the experimental data the pion shadowing is weaker than in
the model and a pronounced flow in ${\rm E_T}$ is the result. 

A first study of the effect of nucleon mean fields on the RQMD
results for proton spectra and flow observables was presented in
\cite{mattiello}. Although the mean field is introduced in a simplified
Skyrme-type parameterization of the interaction, it is obvious that the
model calculations are moving in the right direction. Introducing this
additional repulsion the proton spectra become flatter, the proton flow
increases and the pion shadowing is reduced (see Figures in
\cite{mattiello}). This observation is related to the study of 
the energy dependent trajectories in nn scattering in the ARC
simulations where leaving out the dominantly repulsive character also
drastically reduces the flow prediction.
 
Using the extracted flow parameters for nucleons and employing once more
our knowledge of the proton spectra (see above) we can evaluate $\langle 
p_x \rangle$ as a function of the rapidity and determine the slope
d$\langle p_x \rangle$/dy in
order to compare to data available from lower beam energies. In the
literature, 
typically a slope with respect to rapidity normalized to beam rapidity
is quoted. From the present analysis we find for this slope a value of
d$\langle p_x \rangle$/dy/y$_{b}$ = 0.10 GeV/c. Recently, a systematics
of this variable was shown for beam 
kinetic energies of 0.1 - 2.0 GeV \cite{flow_slope}. In order to
compare different collision systems the slope constants were divided by
the sum of the cube root of target and projectile mass number. It was
observed \cite{flow_slope} that this normalized slope rises with
beam kinetic energy and reaches an approximate plateau in the energy
range of about 0.7 - 2.0 GeV/nucleon with values of 35 - 40 MeV/c. After
normalization to the mass number of the colliding system our present
analysis gives a value for this slope of 35 MeV/c, i.e. practically the
same value as observed at much lower beam energies. This result is unexpected
since the beam momentum and also the proton transverse momenta are much
larger in our case. It is not clear why the slope of the absolute
directed transverse momentum with respect to normalized rapidity should
scale with beam energy. 

\section{CONCLUSIONS}

In summary, the azimuthal distributions of transverse energy and charged
particle multiplity were studied systematically as a function of
pseudorapidity and of centrality for 10.8 A GeV/c Au + Au collisions. A
pronounced dipole component or flow is observed. It crosses zero and
changes sign around mid
rapidity. The magnitude of this flow effect peaks
at intermediate centralities and vanishes for very central
collisions. In addition a much smaller quadrupole component or elliptic
eventshape is observed. The long axis is oriented in the reaction plane
and there is no significant rapidity dependence. 

The same type of
analysis has been performed on events from the generator RQMD and a flow
signal is observed there as well. But it is significantly smaller in the
model than in the data. A different generator, ARC, gives the correct
strength of flow when the NN repulsion is softened at high collision
energies.   

The magnitude of the flow signal is larger in
transverse energy than in charged particle multiplicity and this
difference has been used to extract the flow signal of nucleons and
pions separately for an intermediate centrality bin. It is found that
nucleons show a pronounced flow 
signal while for pions the signal is very weak and in direction opposite
to the nucleon signal. The discrepancy between the data and the model
can be traced to RQMD predicting a weaker proton flow and a stronger 
opposite pion flow as compared to the data. It has been shown in
the literature that introducing a nucleon mean field will improve both
aspects. 

Compared to lower beam 
energies, in the range below 2 GeV kinitic energy per nucleon, the slope
of the directed transverse momentum of protons with respect to
normalized rapidity appears to be about constant while the absolute
rapidity gap between target and projectle and the mean transverse
momentum of protons grow significantly.

\section{ACKNOWLEDGEMENTS}

We thank the AGS staff, W. McGahern and Dr. H. Brown for excellent
support and acknowledge the untiring efforts of R. Hutter in all
technical matters. Financial support from the US DoE, the NSF, the
Canadian NSERC, and CNPq Brazil is gratefully acknowledged. One of us
(JPW) thanks the A. v. Humboldt Foundation for support, while another
(WCC) was supported by the 
Gottlieb Daimler- and Karl Benz-Stiftung for preparation of this manuscript.

\section{FIGURE CAPTIONS}

\begin{itemize}
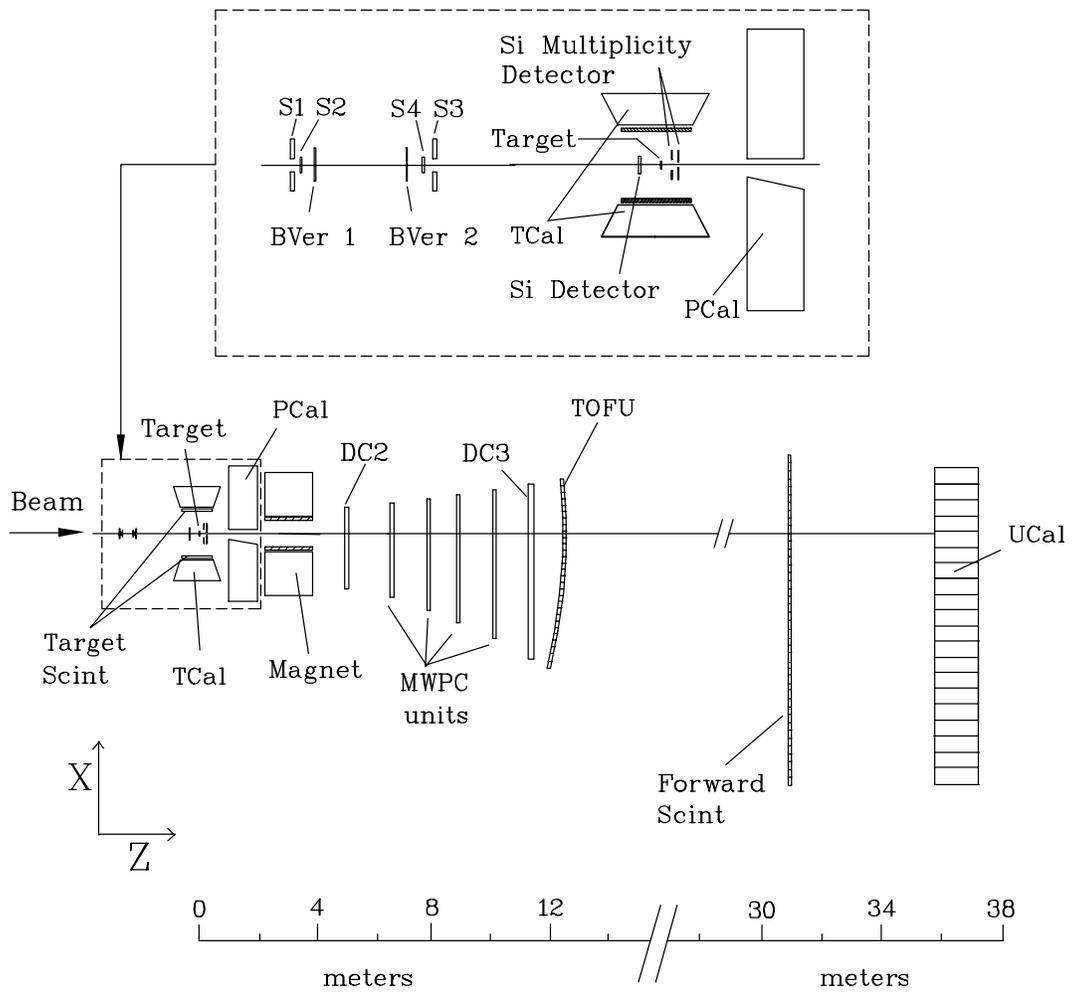
\item[Fig.~1] Experimental setup of E877. The insert shows enlarged
the beam definition and the region surrounding the target.

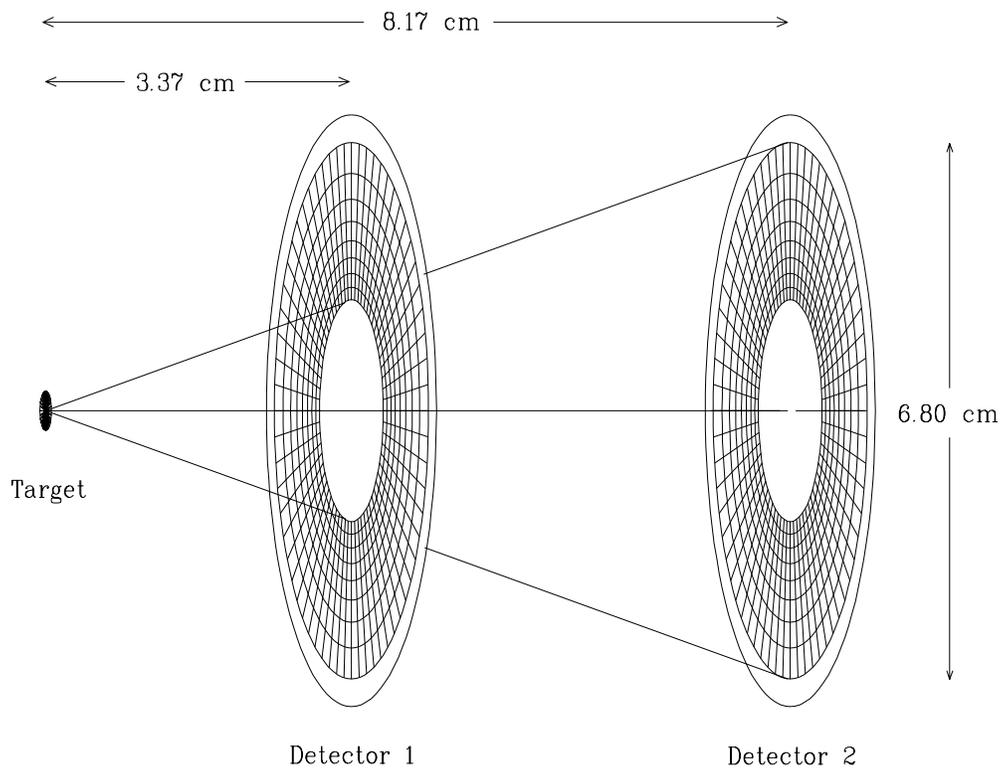
\item[Fig.~2] The participant calorimeter (PCal) and its segmentation,
viewed from downstream. 

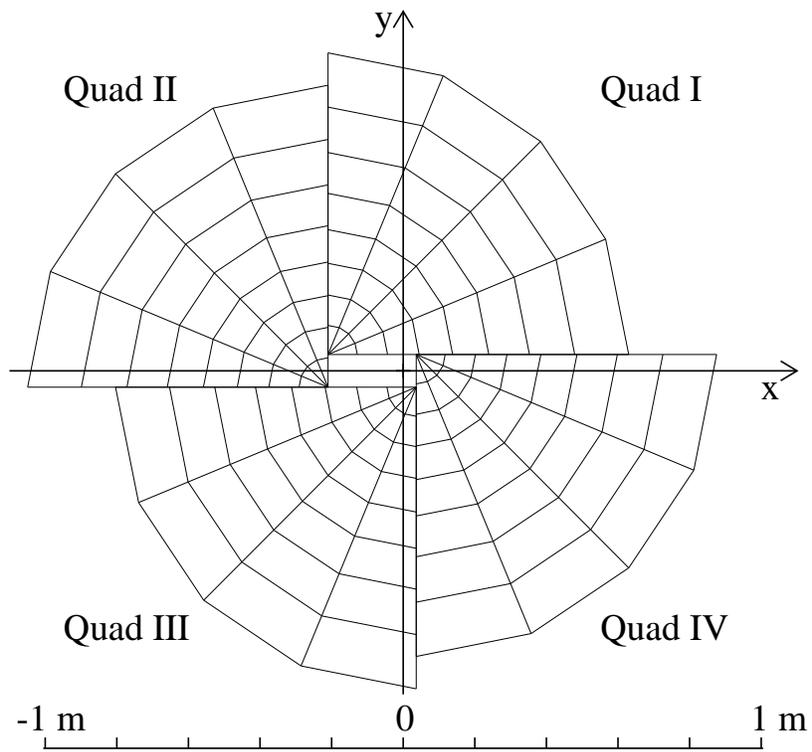
\item[Fig.~3] Placement relative to the target and segmentation of the
silicon pad multiplicity detectors. 

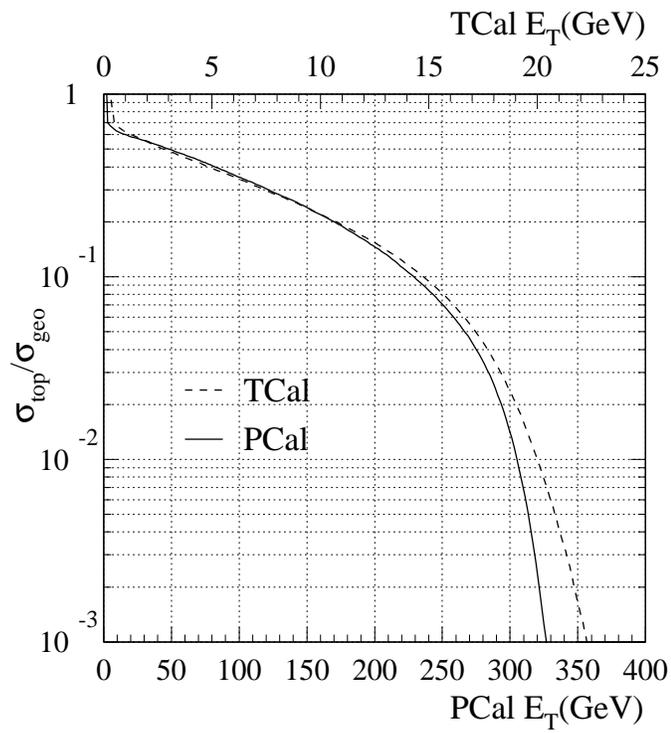
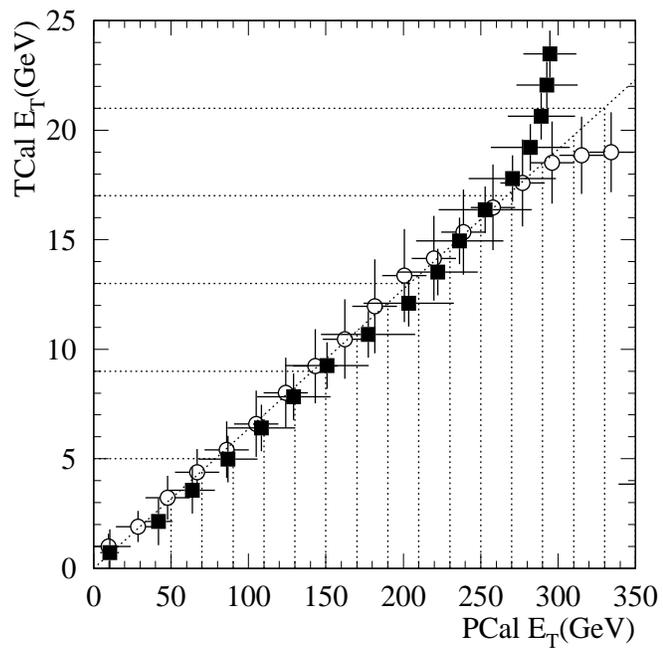
\item[Fig.~4] a) Integral of the measured transverse energy spectrum
for the two calorimeters TCal (dashed) and PCal (solid). The vertical
axis is the cross section integrating from a given ${\rm E_T}$ up to the
top end of the spectrum normalized to the geometric cross section.
b) Correlation of the measured transverse energies in the
two calorimeters projecting on the PCal (open circles) and on the Tcal
(solid squares) scale. The error bars indicate the width (standard
deviation) of the 
correlation. Also indicated are the centrality bins 
(horizontal and vertical dashed lines) used in the analysis (see
text for details).

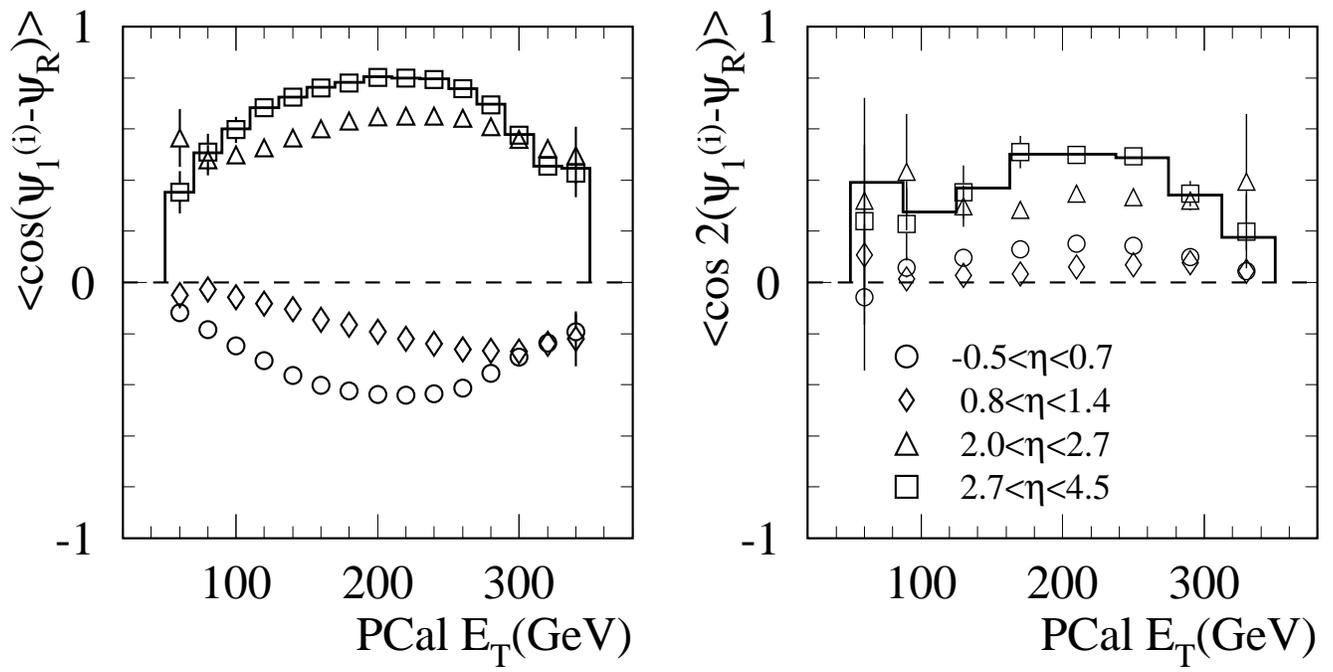
\item[Fig.~5] Inverse correction factor for the first moment $v_1$
(left) and the second moment $v_2$ (right) due to the finite resolution
of the reaction plane angle $\Psi_R$ measurement in four different bins
of pseudo-rapidity (see equation (5)). Solid histogram: correction factors
for the most forward pseudorapidity bin obtained after normalizing to
the 'mixed event' distribution (see text).

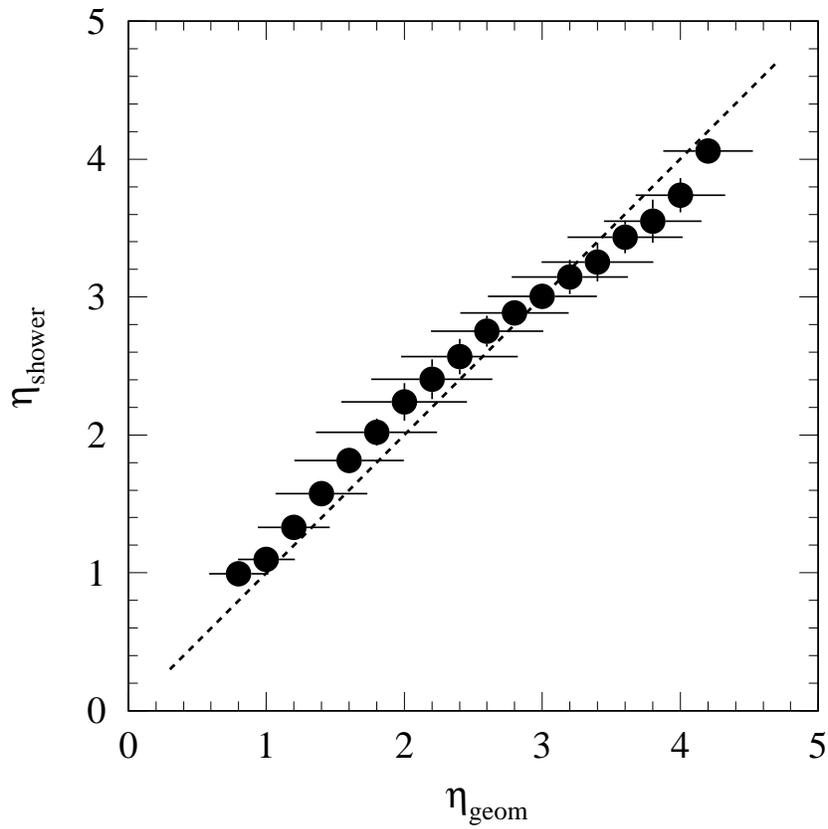
\item[Fig.~6] Pseudorapidity of the particles depositing energy in a
PCal cell as compared to the value corresponding to the geometric center
of each cell. The error bars indicate the range of primary
pseudorapidities contributing to energy deposit in a given cell.

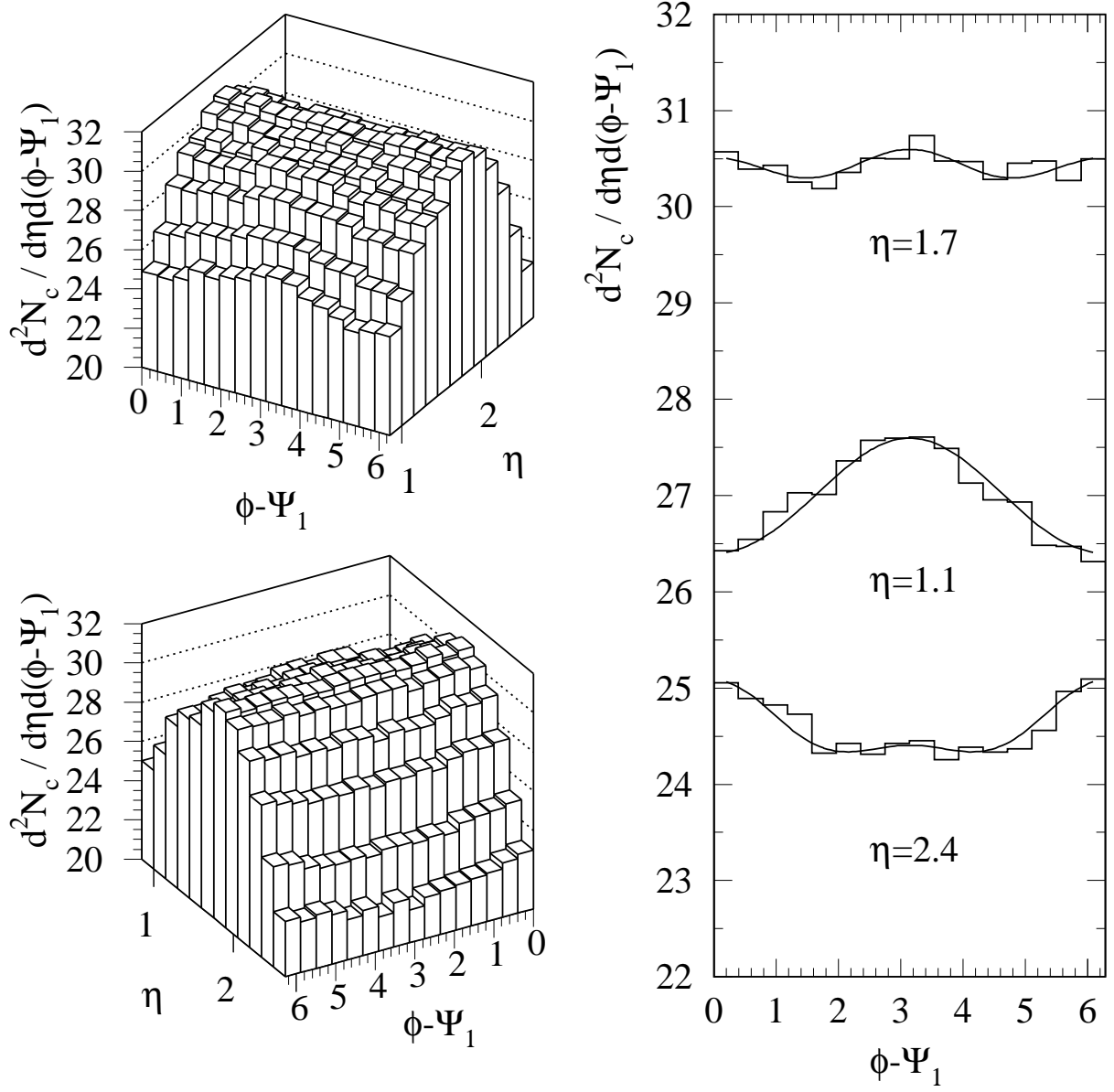
\item[Fig.~7] Left: Double differential charged particle distribution for the
intermediate TCal centrality bin. Right: Three pseudorapidity bins of
the same distribution. The solid line is a distribution with Fourier
coefficients $v_0, v_1, v_2$. 

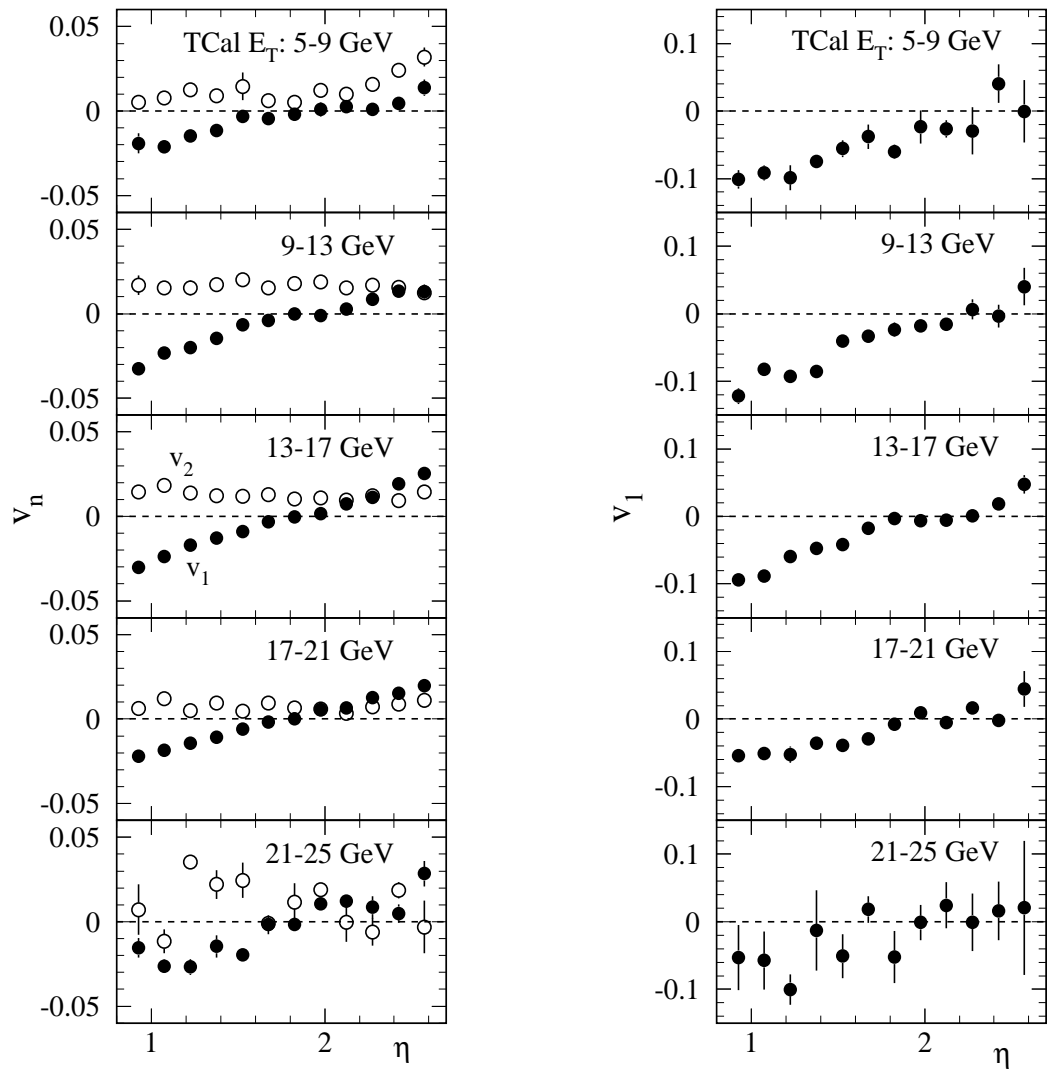
\item[Fig.~8] Left: Flow parameters $v_1$ and $v_2$ for all charged particles.
Right: Flow parameters $v_1$ extracted only for heavily ionizing charged
particles (see text). 

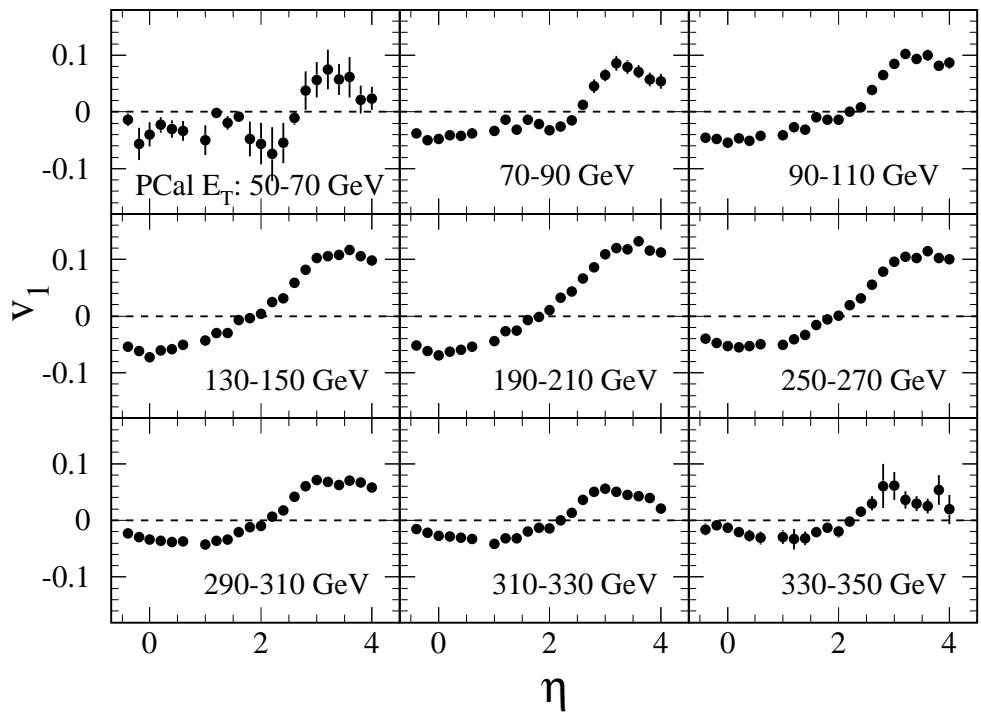
\item[Fig.~9] Flow parameters $v_1$ of the ${\rm E_T}$ azimuthal distribution
after correction of contributions other than from the target for
selected centrality bins (PCal ${\rm E_T}$).

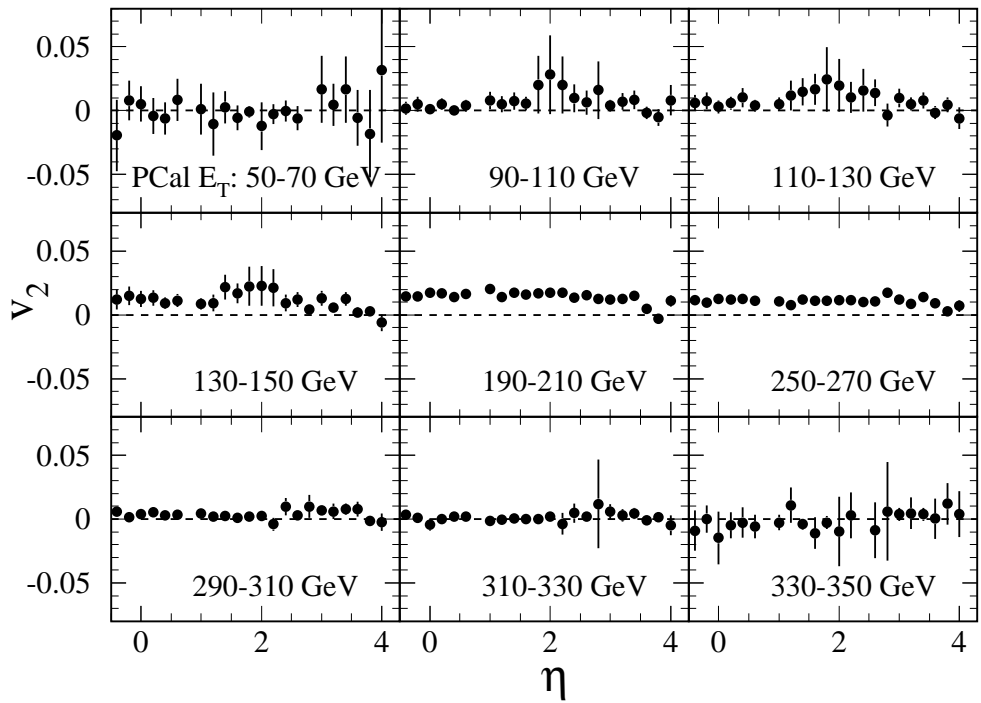
\item[Fig.~10] Flow parameters $v_2$ of the ${\rm E_T}$ azimuthal
distribution for selected centrality bins.

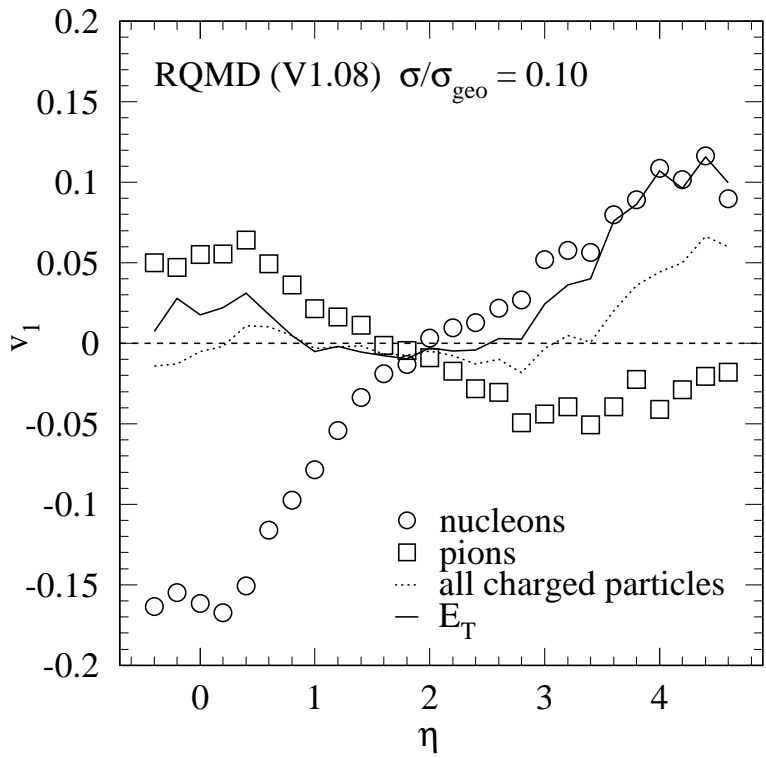
\item[Fig.~11] Comparison of the measured flow parameters $v_1^{({\rm
N_c})}$ and 
$v_1^{({\rm E_T})}$ (solid symbols) for the centrality range 5-15~\%. Also
shown are the equivalent parameters extracted from RQMD events.    

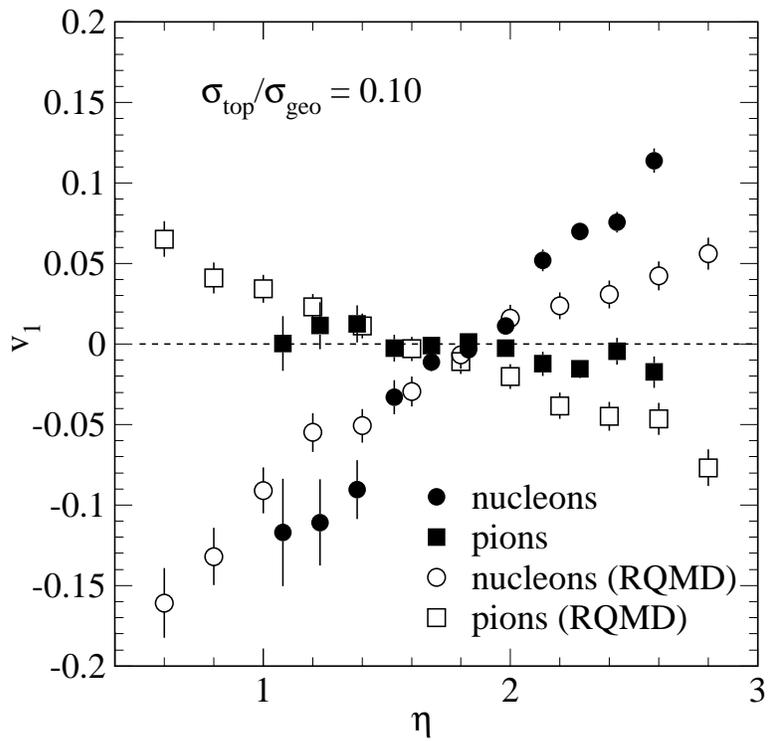
\item[Fig.~12] Flow parameters from RQMD events for nucleons, pions,
transverse energy and charged particle multiplicity for an exclusive
centrality bin ranging from 5-15~\%. The fluctuations are statistical. 

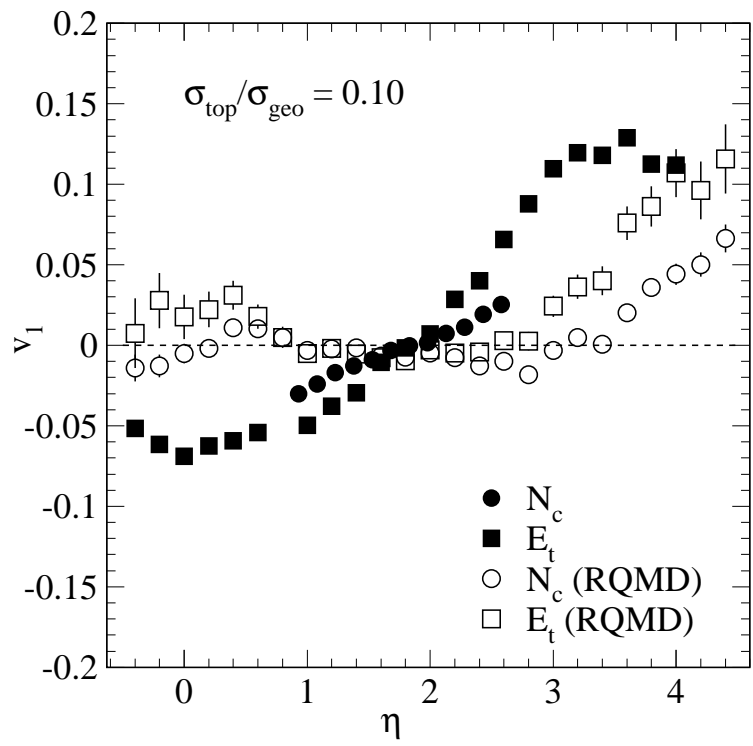
\item[Fig.~13] Decomposition of the flow parameters $v_1^{({\rm E_T})}$ of 
nucleons and pions (solid symbols) and comparison to the extracted
parameters from RQMD (open symbols) for an exclusive centrality bin
ranging from 5-15~\%.

\end{itemize}

\end{document}